\documentclass[12pt]{article}
\usepackage{graphicx}

\usepackage{color}

\newcommand{\beq}{\begin{equation}}
\newcommand{\eeq}{\end{equation}}
\newcommand{\ri}{\mathrm{i}}

\begin{document}

\begin{titlepage}

\vskip 2cm
\begin{center}
{\Large\bf  Quasi-Elastic Neutrino Reactions on Carbon and Lead Nuclei
\footnote{{\tt mh1@ualberta.ca}, {\tt tfinlay1@gmail.com}, {\tt smassoudi@gmail.com}, {\tt cnokes@ualberta.ca}, {\tt mdemonti@ualberta.ca}}}
 \vskip 10pt
{\bf
Mohammad Hedayatipoor$^a$, James Finlay$^{b,c}$, Soheyl Massoudi$^{b,d}$,\\ Charles Nokes$^{b,e}$, Marc de Montigny$^b$}
\vskip 5pt
{\sl $^a$Department of Physics, University of Alberta\\
Edmonton AB, T6G 2E1, Canada}
\vskip 2pt
{\sl $^b$Facult\'e Saint-Jean, University of Alberta \\
 Edmonton, Alberta, Canada T6C 4G9, Canada}
\vskip 2pt
{\sl $^c$Current address: Microsoft Corporation, Corporate Headquarters \\
 One Microsoft Way, Redmond WA, 98052, USA}
\vskip 2pt
{\sl $^d$Current address: \'Ecole polytechnique f\'ed\'erale de Lausanne \\
 Route Cantonale, CH-1015, Lausanne, Switzerland}
 \vskip 2pt
 {\sl $^e$Current address: Department of Physics, University of Alberta\\
Edmonton AB, T6G 2E1, Canada}
\end{center}


\begin{abstract}
We examine neutral-current quasi-elastic neutrino-nucleus reactions on $^{12}$C  and $^{208}$Pb targets. We use the relativistic mean field theory approach to describe the nuclear dynamics. We compute the cross sections for the scattering of 150-MeV,  500-MeV and 1000-MeV neutrinos on a $^{12}$C target  and study the effect of the strange-quark content of the nucleon which appears in these reactions via the isoscalar weak current. We compare our results with the data of the MiniBooNE experiment for mineral oil (CH$_2$).  We also calculate the cross section for the quasi-elastic neutron knockout reaction of 20 to 60-MeV neutrinos on a $^{208}$Pb target which is relevant to plans to use lead as a target material in future supernova neutrino detectors. 
\end{abstract}

\bigskip



\end{titlepage}

\newpage

\setcounter{footnote}{0} \setcounter{page}{1} \setcounter{section}{0} %
\setcounter{subsection}{0} \setcounter{subsubsection}{0}

\newcommand{\anu}{\overline\nu}



\section{Introduction{\label{introduction}}}

The general objective of this paper is to examine the interaction between neutrinos and nuclei by using a model based on the relativistic formalism. We compute the cross sections for the neutrino neutral-current quasi-elastic reaction on carbon and lead targets, with one knocked-out nucleon.

Much of the current research on neutrinos aims at better understanding their intrinsic nature, which may provide evidence for physics beyond the standard model of weak interactions  \cite{Zuber}. Some examples of questions regarding the intrinsic nature of neutrinos include: their Dirac-versus-Majorana type \cite{DiracMajorana}, their  magnetic moment  \cite{magneticmoment},  their role as representatives of CP violation in the leptonic sector \cite{nunokawa}, and their matter-enhanced oscillations \cite{balantekin}.  Since neutrinos are not significantly attenuated when they travel through the interstellar medium, they have applications in astrophysics, such as their contribution to the production of energy in the Sun, the influence of neutrinos on the dynamics of a core-collapse supernova explosion, and the cooling of a proto-neutron star \cite{Bahcall}. In many astrophysical situations, the neutrinos serve as messengers probing the interior of dense and opaque objects that otherwise remain out of reach.

The relativistic approach to nuclear dynamics we use here is based on the relativistic mean field theory developed by Walecka {\it et al} \cite{S}-\cite{SW}. This approach has been applied successfully in the analysis of nucleon knockout reactions using electromagnetic probes and meson photoproduction reactions \cite{EDC.HSS}-\cite{MH.HSS}.  For instance, the model was utilized and successfully compared with experimental data for the reactions $e+X^A\rightarrow e'+N+X^{A-1}$, with $X^A$ oxygen, zirconium and lead,  in a study of relativistic and non-relativistic description of quasi-free electron scattering \cite{MH.JIJ.HSS}. We shall exploit this model and analyze the neutral-current neutrino scattering by two targets: first by $^{12}$C, in order to examine the strange-quark contributions in the nucleon which occur via the isoscalar weak current (and we compare with the data of the MiniBooNE experiment for mineral oil (CH$_2$) \cite{Aguilar-Arevalo}), and second, by $^{208}$Pb in the energy range of neutrinos emitted by a supernova core collapse. 

The neutrino neutral-current scattering allows us to examine the strange quark content of the nucleon, which can manifest itself via the isoscalar weak current. This is different from the charged-current scattering, which involves only isovector weak currents.
The analysis of the strange-quark content of the nucleon originated after suggestions that the measurements of the semileptonic weak-neutral-current reactions $\nu p\rightarrow \nu p$ and ${\overline{\nu}} p\rightarrow {\overline{\nu}}  p$ at Brookhaven National Laboratory (BNL) led to a non-trivial contribution for the strange-quark axial-vector form factor \cite{Ahrens1987}. Further analyses, which also took into account the strange-quark vector form factors, showed that BNL's results could not provide decisive conclusions \cite{Garvey1}-\cite{Garvey3}. The strange-quark contributions to the nucleon form factors were studied with the help of neutrinos by using the relativistic Fermi gas model (RFG) in Refs.\cite{horowitz}-\cite{mulders}. Subsequent calculations based on the relativistic plane-wave impulse approximation (RPWIA) have been compared to RFG calculations in Ref.\cite{alberico1}-\cite{maieron}.  The next step is taking into account the final-state interaction (FSI) of the ejected nucleon \cite{alberico1}-\cite{bleve}.  The relativistic distorted-wave impulse approximation (RDWIA), which we use here, was employed in several neutrino-nucleus calculations  \cite{alberico1}-\cite{maieron}. Ryckebusch {\it et al} performed similar calculations using a relativistic multiple scattering approximation \cite{Belgians1}.  The authors of Ref.\cite{NPA773} utilized the RDWIA for charged- and neutral-current quasi-elastic neutrino-nucleus reactions on $^{12}$C and have discussed the sensitivity of these reactions to the strange-quark content of the nucleon and to the final state interactions (FSI).  

We also extend our computations to the interaction of the neutrinos with $^{208}$Pb. We do this in the energy range which is relevant to the neutrinos resulting from supernova collapse. In this respect there are plans for the Canadian Helium and Lead Observatory (HALO), a detector dedicated to the observation of supernova neutrinos that is located at SNOLAB in the Creighton Mine in Sudbury, Canada \cite{HaloZuber,halowebsite}. It is part of the SuperNova Early Warning System (SNEWS), a worldwide network of detectors currently running or nearing completion, and which are sensitive to core-collapse supernova neutrino signals in the Milky Way. Currently, SNEWS involves seven neutrino experiments \cite{snews}-\cite{snewswebsite}. In addition to providing information about the neutrinos themselves, SNEWS will also allow astrophysicists to learn about the nature of the supernova's core collapse. An important feature of the neutrino signal is that it is prompt: it emerges from the supernova core within tens of seconds, whereas it may take hours or days after the stellar collapse for the electromagnetic signal to emerge. Hence, the neutrinos can provide an early warning to astronomers to observe the very early turn-on of the supernova light curve.  Since lead has a larger neutrino-scattering cross section per nucleon than many other elements, most of the scattering events will produce neutrons, thereby signaling a galactic supernova. Lead has another advantage that, being a double magic nucleus, its neutron-capture cross section is low, so that the neutrons have a higher chance of surviving the trip through the lead and the moderator into the neutron detectors.

In this paper, we examine the neutrino neutral-current reaction on nuclei that results in one knocked-out nucleon $N$:
$X^A(\nu, \nu'N)X^{A-1}$,
where $A$ is the mass number of the nucleus $X$, hereafter, $^{12}$C and $^{208}$Pb. For the reactions with $^{12}$C, $N$ denotes either a proton or a neutron, whereas for $^{208}$Pb, we will consider only $N$ as neutrons.  In Section \ref{formalism}, we describe the neutrino-nucleus scattering model we use. The S-matrix is expressed in terms of lepton and nuclear currents, and we explain how the latter contains the form factors that pertain to the strange-quark parameters that we investigate. We state and discuss our results in Section \ref{results}, for both carbon and lead.


\section{Neutrino-nucleus scattering model {\label{formalism}}}

The relativistic S-matrix, which describes the
quasi-free neutrino scattering process can be written in a manner similar to what is done for quasi-free electron scattering. In the distorted-wave Born approximation (DWBA) the S-matrix for the neutrino reaction takes the form \cite{BD, JU, MH.JIJ.HSS,PRC69},
\begin{eqnarray}
S_{fi} &=& \frac{1 }{(2\pi)^{7/2}}\frac{G_F}{2\sqrt{2}}\delta(E_N+\epsilon_f-\epsilon_i-E_B) 
      \sum_{J_B M_B} { \left(J_f, J_B;M_f, M_B| J_i, M_i \right)}
         \nonumber \\ 
       &  &\times{ \left[ {\cal S}_{J_i J_f} (J_B) \right] }^{1/2} 
                {\cal L}^\alpha N_{\alpha MM_B},
         \label{f1}
\end{eqnarray}
where $G_F$ is Fermi's coupling constant, ${\cal S}_{J_i J_f} (J_B)$ is the spectroscopic factor, $\left| J_iM_i\right\rangle$ and $\left| J_fM_f\right\rangle$ are initial target and final nucleus states, $E_N$ is the energy of the knocked-out nucleon, $\epsilon_i$ and $\epsilon_f$ denote the initial and final energy of  the  neutrino, and $E_B$ is  the  energy of the bound nucleon. The interaction of the neutrino with the nucleus is defined through a relativistic formalism using lepton and nuclear currents ${\cal L}^\alpha$ and $N^\beta_{MM_B}$, respectively. The lepton current is given by
\begin{eqnarray}
{\cal L}^\alpha
        = \overline{\nu}(k_f) (\gamma^\alpha - \gamma^\alpha\gamma^5)\nu(k_i), 
\end{eqnarray}
where $\nu(k_i) $ and $\nu(k_f) $ are the initial and final states of  the  neutrinos, respectively, and $k$ is  the  momentum of  the  neutrino.  We take these neutrino wave-functions as free left-handed Dirac spinors, and $\gamma^\alpha$ and $\gamma^5$ are the well-known Dirac matrices. 
The nuclear current is given by
\begin{eqnarray}
N_{\alpha MM_B}
      = \int d^3x\overline{\psi}_{M}\left( k_p,x \right) j_\alpha
          \psi_{J_BM_B}\left( x \right)e^{\ri{\bf q}\cdot{\bf x}},
          \label{f3}
\end{eqnarray}
where the weak nuclear current operator $j^\mu$ is described below.  
$\psi_{J_BM_B}(x)$ and $\psi_{M}(x)$ are the initial and final states of the nucleons, and $M_B$ and $M$ are their spin projections. These wave-functions are solutions of the Dirac equation with the proper scalar and vector potentials.   The details of the Fock space calculations and the expansion of the Dirac wave-function in partial waves can be found in Ref.\cite{JIJ.HSS}. 

With the appropriate factors of $\hbar$ and $c$, the relativistic expression for the triple differential cross section takes the form
\beq
\frac{d^3\sigma}{d\Omega_{e} d\Omega_p dE_p} =
  \frac{G_F^2 m_N c^2 p_N c k_f^2 c^2} {8\left( 2 \pi \right)^{5}\hbar c} \sum_{ J_B M_B M } 
    { \frac{ {\cal S}_{J_i J_f} (J_B) }{ 2J_B + 1 } }
    { | {\cal L}^\alpha N_{\alpha MM_B} | }^2,\label{crosssection}
 \eeq
We use the maximum value of $2J_B +1$ for spectroscopic factor ${\cal S}_{J_i J_f} (J_B)$.
The Fermi constant is $G_F=1.16639\times 10^{-11}$ MeV$^{-2}$, $k_f$ denotes the final momentum of the neutrino, $p_N$  and $m_N$ are momentum and mass of the emitted nucleon, respectively.

We utilize weak form factors similar to those in Refs.\cite{meucci1}, \cite{NPA773}, and \cite{PRC77}. The authors of Ref.\cite{meucci1} use the following one-particle current operator for the weak current:
\beq
j^\mu = F_1^V(Q^2)\gamma^\mu+\ri\frac\kappa{2M}F_2^V(Q^2)\sigma^{\mu\nu}q_\nu-G_A(Q^2)\gamma^\mu\gamma^5,
\label{FormFactor}\eeq
where $Q^2=|{\bf q}|^2-\omega^2$ is the four-momentum transfer (with four-momentum  $q^\mu= k^{\mu}_{i} - k^{\mu}_{f}$ ), $\kappa$ is the anomalous part of the magnetic moment for nucleon, and  $\sigma^{\mu\nu}=\frac \ri2[\gamma^\mu,\gamma^\nu]$ is the usual commutator of the Dirac matrices. The $Q^2$-dependent functions $F_1^V$ and $F_2^V$ are the weak isovector Dirac and Pauli form factors, respectively, and they are given by
\beq
F_i^{V,p(n)}=\left(\frac 12-2\sin^2\theta_W\right)F_i^{p(n)}-\frac 12F_i^{n(p)}-\frac 12F_i^{s}, \quad i=1, 2,\label{FiV}
\eeq
where we have taken the Weinberg angle $\theta_W$ as  $\sin^2\theta_W\simeq0.23143$, and the electromagnetic form factors  $F_i^p$ and $F_i^n$ are as in Ref.\cite{3WW-DP}.
In Eq.(\ref{FiV}), $F_1^s$ and $F_2^s$ are the strangeness contributions to the vector form factors \cite{PRC77,strange},
\beq
F_1^s(Q^2)=\frac{(\rho^s+\mu^s)\tau}{(1+\tau)(1+Q^2/M_V^2)^2}, \qquad F_2^s(Q^2)=\frac{(\mu^s-\tau\rho^s)}{(1+\tau)(1+Q^2/M_V^2)^2}\label{FsItalians}
\eeq
where $\tau=Q^2/(4m_N^2)$, $M_V=0.843$ GeV, and the strangeness parameters $\rho^s$ and $\mu^s$ will be given various values, as described in Section \ref{results}. 
The remaining element in Eq.(\ref{FormFactor}) is the function $G_A$, which denotes the axial form factor \cite{axial},
\beq
G_A(Q^2)=\frac 12(\tau_3g_A-g_A^s)G(Q^2),\label{gAs}
\eeq
where $g_A\simeq 1.26$,  $G=(1+Q^2/M_A^2)^{-2}$, with $M_A=(1.026\pm 0.021)$ GeV, and $\tau_3=+1/-1$ for proton/neutron knockout reactions. The parameter $g_A^s$ describes various strange-quark contributions and, like $\rho^s$ and $\mu^s$ in Eq.(\ref{FsItalians}), we will consider different values in Section \ref{results}.


\section{Results and analysis {\label{results}}}

In Section \ref{carbon}, we  discuss the results of our computations of the cross section for the quasi-elastic scattering of neutrinos on a $^{12}$C target, with neutrino energy equal to 150, 500 and 1000 MeV. In these calculations we study the effects of the strange-quark contributions to nucleons. We also compare the result of the model for mineral oil with the data available from the MiniBooNE experiment \cite{Aguilar-Arevalo}. In Section \ref{lead}, we discuss the scattering cross section of neutrinos on a $^{208}$Pb target. The neutrinos in this reaction have a lower energy range, 20 - 60 MeV, which is relevant to plans to use lead as a target in future supernova neutrino detectors. 

In all the calculations presented in this work the symmetrized Wood-Saxon potentials are used for bound state wave-functions \cite{4GER.THS}.
The continuum wave-functions for the knocked out nucleon are obtained
using the energy and the $A$-dependent optical potential of Cooper {\it et al} \cite{COPE}.

\subsection{Scattering from carbon:  strange-quark content of the nucleon and scattering on CH$_2$ {\label{carbon}}}

The purpose of this section is twofold. In the following subsection \ref{strange}, we examine the effect of the strangeness parameters on the neutrino-nucleus reactions. We compute the cross sections for neutral-current quasi-elastic  neutrino scattering from $^{12}$C for various values of the strangeness parameters $\rho^s$, $\mu^s$, $g^s_A$  (Eqs.\ref{FsItalians} and \ref{gAs}). This is done for neutrino energies of 150, 500 and 1000 MeV. We examine the strange-quark contributions to the form factors and compare our cross sections with the literature. 

In the subsection \ref{cdata}, we compare our results with data from the MiniBooNE experiment for mineral oil (CH$_2$) target \cite{Aguilar-Arevalo}. Their data result from a high-statistics measurement of the flux-averaged cross section as a function of the momentum transfer $Q^2$  for $Q^2<1.6$ GeV$^2$  \cite{Aguilar-Arevalo}.

\subsubsection{Effect of the strange-quark contribution on neutrino-nucleus reactions.}{\label{strange}}

Before we compare our results with the experimental data for the quasi-free neutrino scattering we will do an initial comparison with the previous work for the role of the strangeness parameters.  In Figs.\ref{n150} to \ref{rpn1000}, we show the dependence of the differential cross sections for the reaction on $^{12}$C on the various combinations of the strangeness parameters. We display the differential cross sections for the reaction on $^{12}$C in Figs.\ref{n150} to \ref{p1000} with neutron or proton knockouts. We use the values of the parameters used in reference \cite{NPA773}. 

Figs.\ref{n150} and \ref{p150} show the results for neutron and proton knockout, respectively, for 150-MeV neutrinos. They display the differential cross section versus the kinetic energy of the knocked-out nucleon. We note that the shapes and magnitudes of the cross sections are similar for proton and neutron, but they differ when it comes to the dependence on the strangeness parameters. For neutron knockout with all the strangeness parameters equal to zero, the cross section is largest, whereas it is smallest for proton. The results presented in these figures indicate that the dependence of the cross section on the strangeness parameter $\rho^s$ is weak. This can be seen by the overlap of the curves of $\mu^s=-0.5$, $g^s_A=-0.1$, with $\rho^s=2$ and $0$. This weaker dependence on $\rho^s$ is echoed by the curves with $\mu^s=0$ and $g^s_A=-0.1$.
 
The results for $E_\nu=500$ MeV are presented in Figs.\ref{n500} for neutrons and Fig.\ref{p500} for protons. These figures indicate that the strangeness parameter $\rho^s$ plays a stronger role for neutrons compared to 150 MeV. However, the dependence on $\rho^s$ continues to be weak for protons.  

Figs.\ref{n1000} and \ref{p1000} show the cross sections for neutron and protons, respectively for $E_\nu=1000$ MeV. From Fig.\ref{n1000} we see that the role of strangeness parameter $\rho^s$ becomes relatively stronger with energy of neutrino.  In the proton case however, the effect of $\rho^s$ continues to be negligible.  We notice from Figs. \ref{n150} to \ref{p1000} that the cross section increases as we lower $\mu_s$ from $0.0$ to $-0.5$. 

When we compare our results with those of Fig.3 of Ref.\cite{NPA773}, we  observe that for $E_\nu = 500$ MeV the general behaviour of our results (in our Figs.\ref {n500} and \ref{p500}) is similar to but slightly smaller than the results in Fig.3 of Ref.\cite{NPA773}. Note that our results display a slight shoulder around $T=200$ MeV, as in Ref.\cite{NPA773}. For $E_\nu =  500$ MeV, the relative curves (with different strangeness values) seem generally ordered in the same way in our results as in Ref.\cite{NPA773}. 

The Fig.1 of Ref.\cite{KimCheoun2008} has some similarities with our results: for knocked-out neutrons, their cross sections for $g_A^s=0.0$ is greater than for $g_A^s=-0.19$ with values similar to ours, and for knocked-out protons,  the contributions are reversed: their cross sections for $g_A^s=0.0$ is smaller than for $g_A^s=-0.19$. 

Figs.\ref{rpn150},  \ref{rpn500} and  \ref{rpn1000} show the proton-to-neutron ratio of cross sections as a function of the knocked-out nucleon kinetic energy for neutrino energies of 150, 500 and 1000 MeV, respectively.  They all show the curves in the same order as Fig.5 of Ref.\cite{NPA773}.  A common feature of these figures is the grouping of curves according to their strangeness:  (1) the zero-strangeness curve is isolated, (2) the two curves with $g^s_A=-0.19$ and  (3) the four curves with $g^s_A=-0.1$, with the cross section increasing with $\left|g^s_A\right|$.  For $E_\nu=500$ MeV, our curves are slightly concave upward and their values are between 0.5 and 0.7.   For $E_\nu=1000$ MeV, our values are smaller than in Fig.5 of Ref.\cite{NPA773}. Our curve for $\rho^s$, $\mu^s$, and $g^s_A$ =  2.0, 0.0,$ -0.10$ crosses 0.0, 0.0, 0.0. That behaviour is not shown in Fig.5 of Ref.\cite{NPA773}.   For strangeness values $\rho^s$, $\mu^s$, and $g^s_A$ = 0.0, 0.0, $-0.10$, the curve in Fig.3 of  Ref.\cite{PRC76} is locally parallel to that with zero strangeness factors, but higher by about 0.1; that is, it lies between 0.9 and 0.95. We observe exactly the same behaviour with our results, although our respective results are slightly shifted downward. 

Fig.4 of Ref.\cite{PRC76} corresponds to $E_\nu=$1000 MeV. There is a big shift between our curves for $\rho^s$, $\mu^s$, and $g^s_A$ equal to 0.4, $-0.31$, $-0.19$ and the other strangeness values.  Three curves can be compared with Fig.4 of Ref.\cite{PRC76}:  $\rho^s$, $\mu^s$, and $g^s_A$ = 0.0, 0.0, 0.0;  0.0, 0.0, $-0.19$; and 0.4, $-0.31$, $-0.19$. The ordering is the same as ours, except for 0.4, $-0.31$, $-0.19$.  The Fig.2 of Ref.\cite{KimCheoun2008} shows the ratios with solid lines for $g_A^s=-0.19$ and dashed lines for $g_A^s=0.0$.  We observe the same ordering in our figure \ref{rpn500}, but the values are slightly different.  In both cases, our proton-to-neutron cross sections are smaller than theirs.

\subsubsection{Comparison with data from the MiniBooNE experiment.}{\label{cdata}  The neutral-current quasi-elastic scattering on CH$_2$ involves three scattering processes: on free protons in hydrogen, bound protons in carbon and bound neutrons in carbon \cite{Aguilar-Arevalo}. The neutrino flux for different types of neutrino species is given in Ref.\cite{Aguilar-Arevalo2}, and we used it to compare our calculations with experiment for the flux-averaged neutrino cross section.  Expressions for flux-averaged  and flux-integrated cross sections are given in Ref.\cite{Butkevich}.
A rather detailed comparison of neutral-current quasi-elastic processes using various nuclear models with the MiniBooNE experiment is available in Ref.\cite{Ivanov}.

The flux-averaged cross section is obtained from the integration over the energy $E_\nu$ of the incoming neutrino:
\beq
\left\langle\frac{d\sigma}{dQ^2}\right\rangle=\int w\left(E_\nu\right)\frac{d\sigma}{dQ^2}\left(E_\nu\right)\; dE_\nu,
\label{flux-averagedcc}\eeq
where the momentum transfer is related to the kinetic energy of the emitted nucleon by $Q^2=2m_NT_N$. The neutrino weight function $w\left(E_\nu\right)=\frac{\phi_\nu\left(E_\nu\right)}{\Phi}$ is defined in terms of the neutrino spectrum of the flux, $\phi_\nu\left(E_\nu\right)$, and the total flux $\Phi$ is,
\beq
\Phi=\int \phi_\nu\left(E_\nu\right)\; dE_\nu.
\eeq
We applied Eq.(\ref{flux-averagedcc}) by computing $\frac{d\sigma}{dQ^2}$, for neutrino energies ranging from 25 MeV to 2975 MeV by interval of $dE_\nu=50$ MeV. We then multiplied each cross section by the weighted flux $w(E_\nu)$ given above and integrated over the neutrino energies.
Comparison of our results with the data from MiniBooNE requires that we compute the following cross section per nucleon:
\beq
\frac{d\sigma_{\nu N\rightarrow\nu N}}{dQ^2}=\frac 17C_{\nu p,H}\frac{d\sigma_{\nu p\rightarrow\nu p,H}}{dQ^2}+\frac 37C_{\nu p,C}\frac{d\sigma_{\nu p\rightarrow\nu p,C}}{dQ^2}+\frac 37C_{\nu n,C}\frac{d\sigma_{\nu n\rightarrow\nu n,C}}{dQ^2},
\eeq
where $C_{\nu p,H}$, $C_{\nu p,C}$ and $C_{\nu n,C}$ are $Q^2$-dependent efficiency correction functions for the neutrino scattering off free proton in H, the bound protons in C and the bound neutrons in C, respectively. The theoretical cross sections $\frac{d\sigma_{\nu p\rightarrow\nu p,H}}{dQ^2}$, $\frac{d\sigma_{\nu p\rightarrow\nu p,C}}{dQ^2}$ and $\frac{d\sigma_{\nu n\rightarrow\nu n,C}}{dQ^2}$ correspond to neutrinos on free protons (per free proton), on bound proton (per bound proton) and on bound neutron (per bound neutron), respectively.

The results are shown in Figs.\ref{square} and \ref{squarelog} along with the MiniBooNE data from Ref.\cite{Aguilar-Arevalo}. 
We observe that whereas the cross section obtained with the plane-wave approximation with no strangeness produces the data quite well, the distorted-wave cross section underestimates the data in the low-$Q^2$ region up to about 0.7 GeV$^2$. The results of the distorted-wave calculations with no strangeness (long dashed curves) is similar to the relativistic mean field (RMF) calculations presented in Fig.3 of Ref.\cite{Ivanov}. The short dashed curve in these figures provides an interesting insight into the effects of the strangeness parameters on the cross sections. The calculations for this curve used a set of the strangeness parameters ($\rho^s=2$, $\mu^s=-0.5$, and $g^s_A-0.1$) discussed in previous subsection which are similar to those used in Ref.\cite{NPA773}. With this set the cross section increases and curve moves up towards the data. This offer a glimpse of hope with improved calculations one might be able to determine the strangeness contribution. Our model is limited to quasi-free one nucleon knockout reaction and lacks from a reliable optical potential for outgoing neutrons.  The behaviour of the cross section at higher momentum transfer is shown with a logarithmic scale in Fig.\ref{squarelog} and shows clearly that the plane-wave calculations fits the data for the complete range and that the distorted-wave calculation also lies within the error bars of the data for $Q^2>0.7$ GeV$^2$.

\subsection{Scattering from lead and supernova core-collapse neutrinos {\label{lead}}}

In this section, we turn to the reactions induced by core-collapse supernova neutrinos, in view of a lead-based observatory such as HALO in SNO+  \cite{HaloZuber,halowebsite}. We apply the formalism discussed above to calculate the cross section for the interaction of these neutrinos with a lead target. 

Kolbe and Langanke computed the cross sections and branching ratios for neutrino-induced reactions for the two materials, lead and iron, for various supernova neutrino spectra. This was motivated by proposed supernova-neutrino and neutrino-oscillation detectors such as MINOS, LAND, OMNIS, which considered one of these materials as target \cite{Kolbe}. Table VI of Ref.\cite{Kolbe} lists the cross section for $(\nu, \nu')$ scattering on lead for incoming neutrino energies between 10 and 150 MeV. 

There are excellent reviews on the processes of core-collapse supernovae; see Refs.\cite{Scholberg, Janka, Kotake} and references therein. Our main interest is the neutrino flux predictions in the neutrinosphere; that is, the surface of last scattering of supernova-emitted neutrinos (see Fig.1 of Ref.\cite{Scholberg}, and Ref.\cite{Vaananen}). Based on various analyses of supernova neutrino spectra, core-collapse supernova leads to a neutrino fluence, or time-integrated flux, $\frac{dF_\nu(E)}{dE}$, which at Earth spreads over an energy range of approximately 10 to 60 MeV  \cite{Beacom, Belina}.  

Whereas at intermediate energy the quasi-elastic knockout is the main contributor to the cross section, at low energy, the quasi-elastic knockout is a contribution that needs to be added to the cross section obtained based on intermediate excited states of the nucleus; the latter were computed in Refs. \cite{Kolbe} and \cite{Engel2003}.

In Fig.\ref{40MeVAll}, we display the differential cross section of the neutron-knockout reaction, $^{208}$Pb$(\nu,\nu'n)^{207}$Pb, computed with the relativistic plane-wave impulse approximation, for the individual energy levels $2d_{3/2}$,  $2d_{5/2}$,  $2f_{5/2}$,  $2f_{7/2}$,  $3p_{3/2}$,  $3p_{1/2}$,  $3s_{1/2}$. The cross sections computed from Eq.(\ref{crosssection}) for those neutrino energies, and considering all three neutrino flavours, are displayed in Table \ref{Table}. It is interesting to observe that our results, based on the quasi-elastic neutron knockout, are comparable to those in Refs.\cite{Kolbe} and \cite{Engel2003}. Note, however, that due to the lack of reliability of the optical potential for lead at low energies, we did not present the distorted-wave calculations, as in the previous section, and we performed plane-wave calculations. Since we expect the effects of distortion to reduce the cross section, the results obtained with the distorted-wave computations should produce smaller values.

\begin{table}
\begin{center}
  \begin{tabular}{ | c |  c | c| }
    \hline
    $E_\nu$ &  $\sigma_{PW}$  &  $\sigma_{1n}$ (Ref.\cite{Engel2003})  \\ \hline\hline
    10 & 0.00 & 0.02 \\ \hline 
    15 &  0.411 &  0.6 \\ \hline
    20 &  2.27 &  2.0 \\ \hline
    25 &  7.05 &  4.6 \\ \hline
    30 &  16.3 &  8.7 \\ \hline
    35 &  30.9 &  14.4 \\ \hline
    40 &  52.2 & 21.5 \\ \hline
    45 &  81.3 &  29.7\\ \hline
    50 &  119 &  38.6 \\ \hline
    55 &  167 &  47.9 \\ \hline
    60 &  224 &  57.4 \\
    \hline
  \end{tabular}
\caption{Total cross section (in units of 10$^{-45}$ m$^2$) of the neutral-current neutrino quasi-elastic scattering on $^{208}$Pb with neutron knockout for various energies in MeV of the incoming neutrino. $\sigma_{PW}$ is computed by using the relativistic plane-wave impulse approximation, and the last column shows the results for $\nu\rightarrow\nu$ from Table 1 in Ref.\cite{Engel2003}.
\label{Table}}
\end{center}
\end{table}

In order to find the number of neutrons created via neutral-current reactions with electron-neutrinos at HALO-1, which consists of 79 tonnes of lead, we multiplied the cross section $\sigma(E)$ by the fluence $\frac{dF_\nu(E)}{dE}$, which produces the event distribution displayed in Fig.\ref{event_distribution}. We use Eq.(3) of Ref.\cite{Beacom} to compute the fluence $\frac{dF_\nu(E)}{dE}$, of each neutrino flavour $\nu$,
\beq
\frac{dF_\nu(E)}{dE}=\left(2.35\times 10^{13}\right) \frac{{\cal E}_\nu}{d^2} \frac{E^3}{\langle E_\nu\rangle^5}\exp\left(-\frac{4E}{\langle E_\nu\rangle}\right),\qquad \left[{\mathrm{in}\ }\frac 1{{\mathrm{cm^2\ MeV}}}\right]
\label{fluence}\eeq
where ${\cal E}_\nu$ is the total energy emitted by the supernova, in units of $10^{52}$ erg, $d$ is the distance between the emitting supernova and the Earth, in unit of 10 kpc, $E$ is the neutrino energy, and $\langle E_\nu\rangle=12$ MeV for the electron neutrino $\nu_e$ considered here.  The fluence distribution for the supernova neutrinos is shown in Fig.\ref{total1t} with the parameters $d=1$ (that is, a distance of 10 kpc from the emitting supernova to the Earth) and ${\cal E}_\nu=5$ (or $5\times10^{52}$ erg)  in Eq.\ref{fluence}. From Fig.\ref{total1t}, we see that the fluence peaks at about $1.7\times 10^{10}$ 1/MeV$\cdot$cm$^2$ around $E_\nu$=8-9 MeV and has decreased by one order of magnitude at $E_\nu$=25 MeV.

The total number of events is obtained by utilizing the flux-integrated neutrino cross section, given the span of energies of the neutrinos emitted by a supernova,
\beq
\left\langle n_{\mbox{event}}\right\rangle=\int dE\; \sigma(E)\;\frac{dF_\nu(E)}{dE},
\label{fluxaveragedcc}\eeq
with $\frac{dF_\nu(E)}{dE}$ from Eq.(\ref{fluence}), and $ \sigma_{PW}(E)$ from Table \ref{Table}. This gives a total of 0.54 events for all three neutrino flavours.  As mentioned earlier, this quasi-elastic one-neutron-knockout contribution should be added to the neutral-current one-neutron-knockout cross section, based on intermediate excited states, utilized in Ref.\cite{Engel2003}. Moreover, the quasi-elastic knockout contributions could be added respectively for multi-neutron knockout reactions as well as charged-current processes. 

In order to get a sense of the number of neutrons produced by HALO-1, we used the cross sections of Table I in Ref.\cite{Engel2003}, computed the probability of neutron production for each flavour, and integrated this probability with the simplified relation for supernova fluence given by Eq.(3) of Ref.\cite{Beacom}. We chose the same values for the fluence parameters as those suggested in Ref.\cite{Beacom}.
This led to 30 neutron events created at HALO-1. The addition of quasi-elastic one-neutron-knockout contribution for the processes will increase the number of neutrons produced at HALO-1.


\section{Conclusions}

In this paper we have presented calculations for the quasi-free scattering of neutrinos in the framework of a relativistic approach. Our focus was on the contributions from the neutral weak current leading to the knockout of a nucleon from the target nucleus. For scattering on carbon, both RDWIA and RPWIA calculations were compared with data from the MiniBooNE experiment. The results obtained with the plane-wave calculations lie between the error bars for the whole range of $Q^2$ while those of the more realistic distorted-wave calculations move below the data for $Q^2$ less than 0.7 GeV$^2$.  We performed calculations using one set of non-trivial strangeness parameters, and this improved the results of the distorted-wave calculations with respect to the data. This is encouraging, and suggests that future improvements to the current model should include strangeness contributions.  

We made an attempt to explore the role of a quasi-free contribution to the cross section on lead. We could only assess that using RPWIA calculations. We observed that the cross sections thus obtained were comparable to those based on intermediate excited states of the nucleus, to which our results need to be added in order to obtain the number of neutrons produced at HALO-1.




\section*{Acknowledgement}

MdM is grateful to the Natural Science and Engineering Research Council (NSERC) of Canada for partial financial support. 
  We are grateful to Helmy S. Sherif for his input and enlightening discussions, and to Kai Zuber and Aksel Hallin for helpful comments.

\newpage

\begin{figure}
\begin{center}
\includegraphics[scale=1.2]{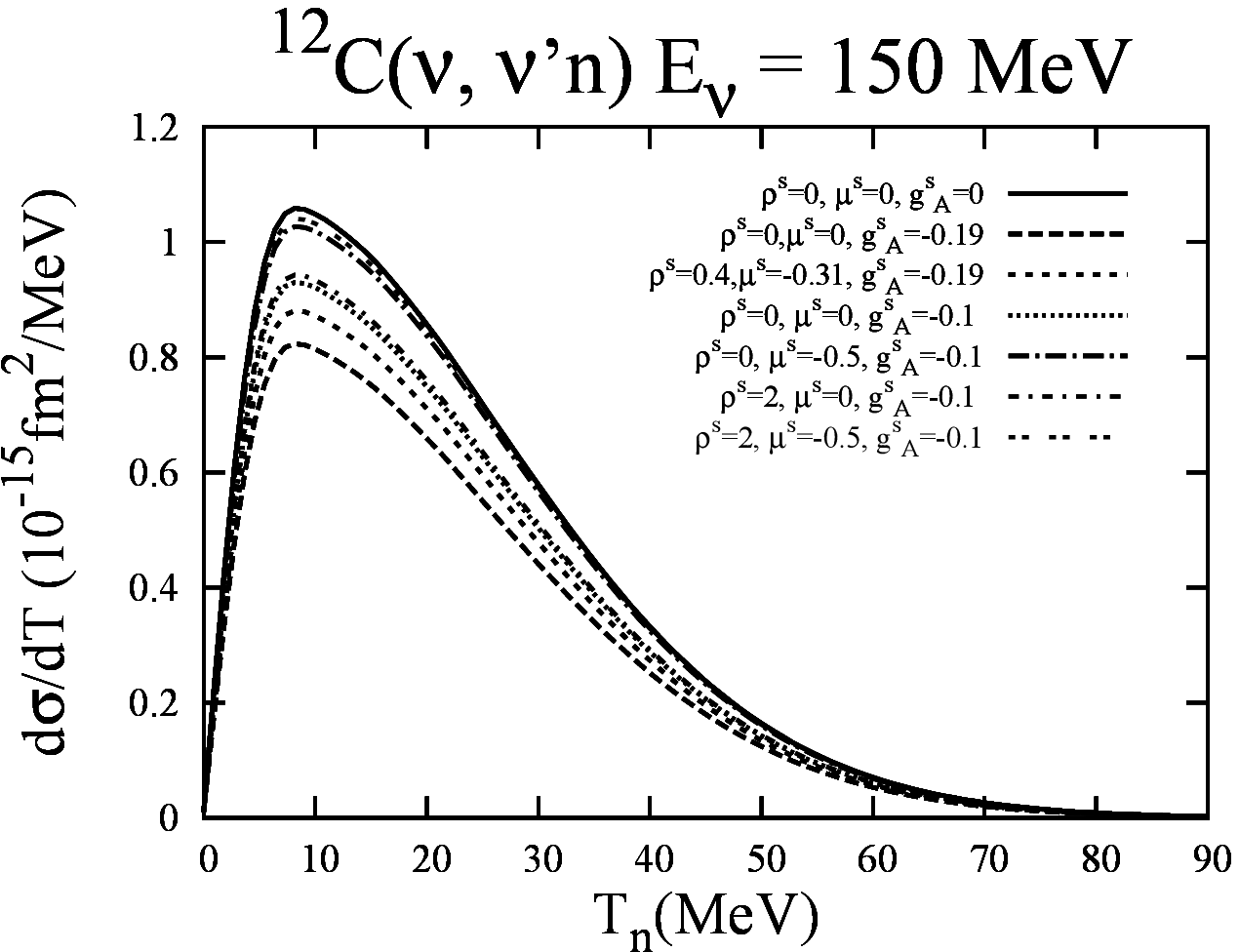}
\end{center}
\caption{Differential cross section of the neutral-current neutrino quasi-elastic scattering on $^{12}$C in terms of the knocked-out neutron kinetic energy for an incoming neutrino energy equal to 150 MeV. The various lines correspond to different strangeness contributions.\label{n150}}
\end{figure}

 \begin{figure}
  \begin{center}
\includegraphics[scale=1.2]{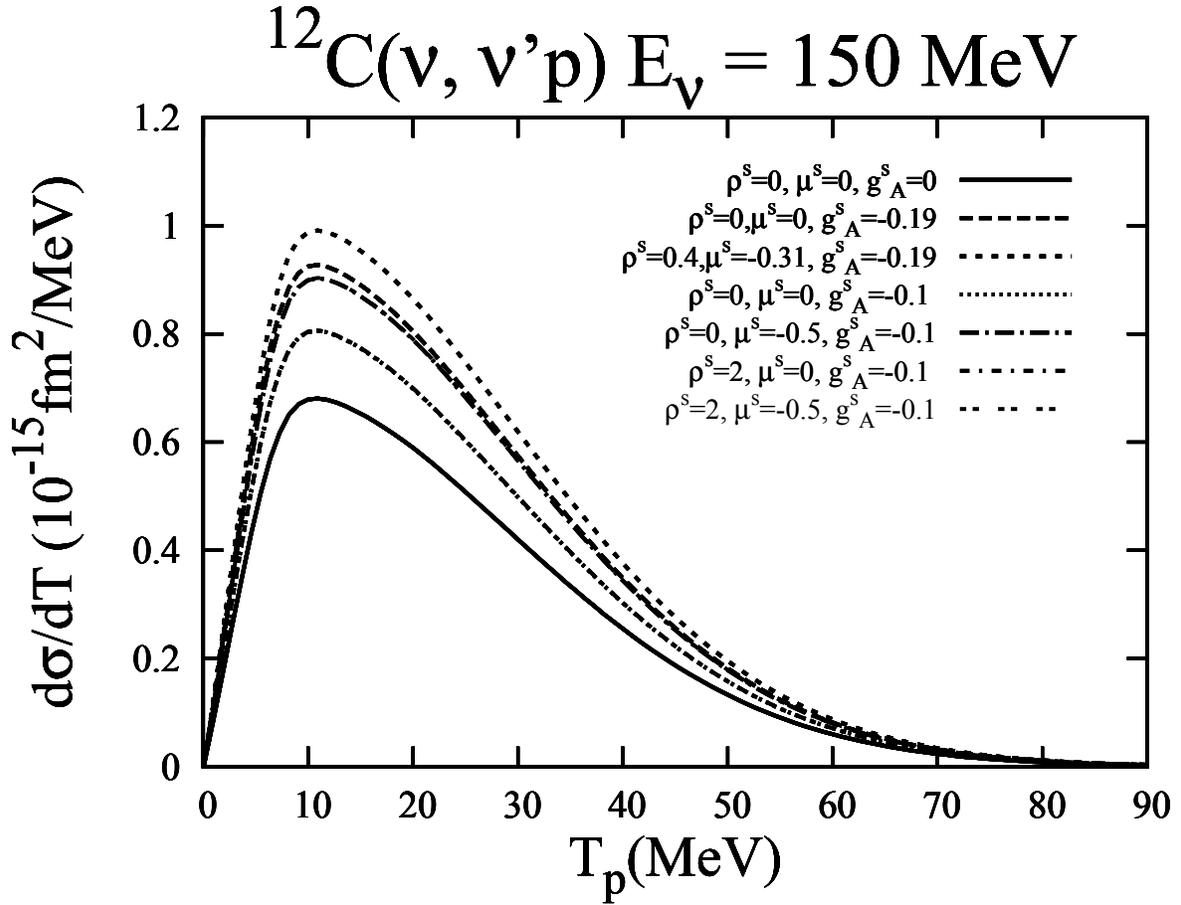}
\end{center}
\caption{Differential cross section of the neutral-current neutrino quasi-elastic scattering on $^{12}$C in terms of the knocked-out proton kinetic energy for an incoming neutrino energy equal to 150 MeV. \label{p150}}
\end{figure}

\begin{figure}
 \begin{center}
  \includegraphics[scale=1.2]{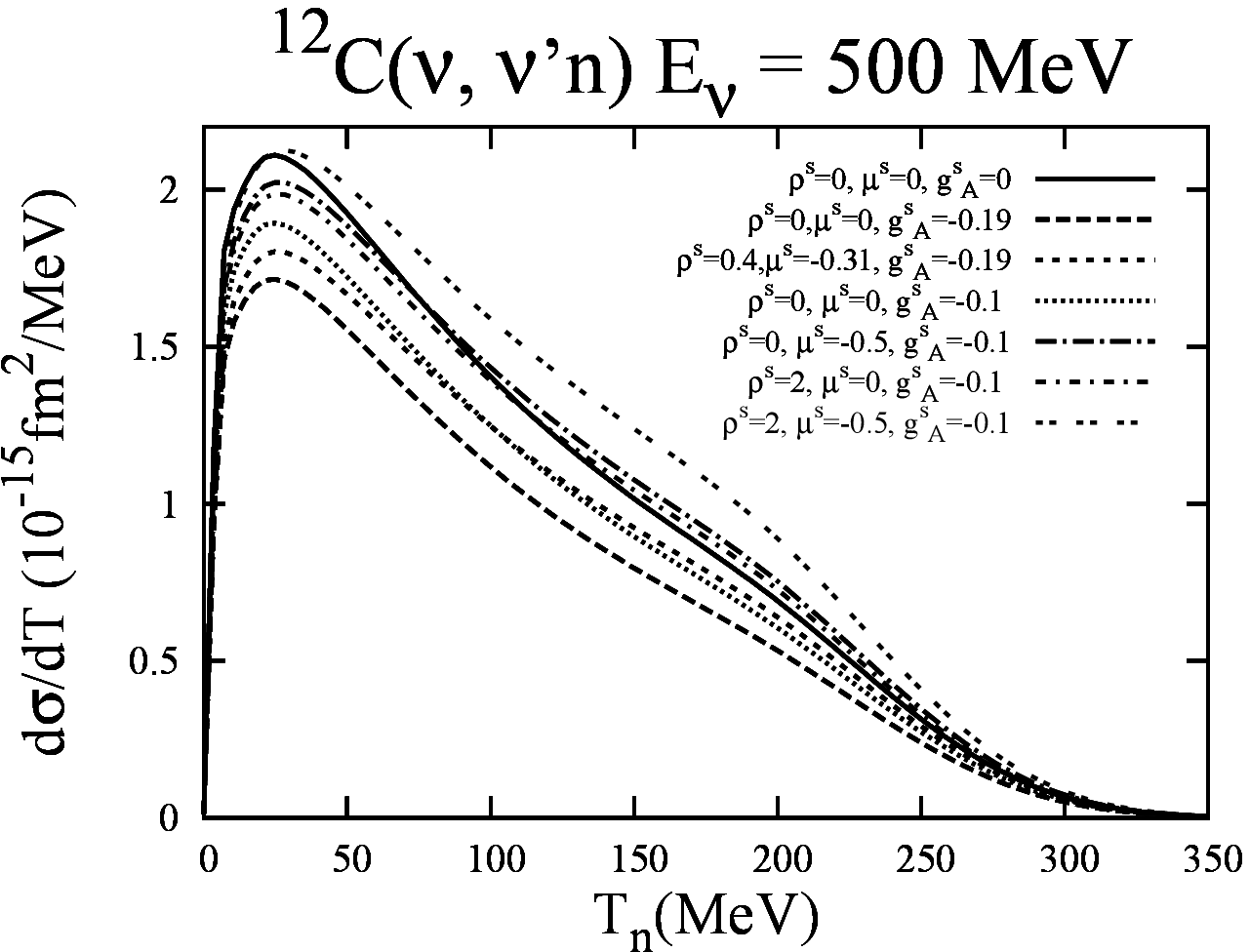}
 \end{center}
 \caption{Same as Figure \ref{n150} for an incoming neutrino energy equal to 500 MeV.\label{n500}}
\end{figure}
 
\begin{figure}
  \begin{center}
  \includegraphics[scale=1.2]{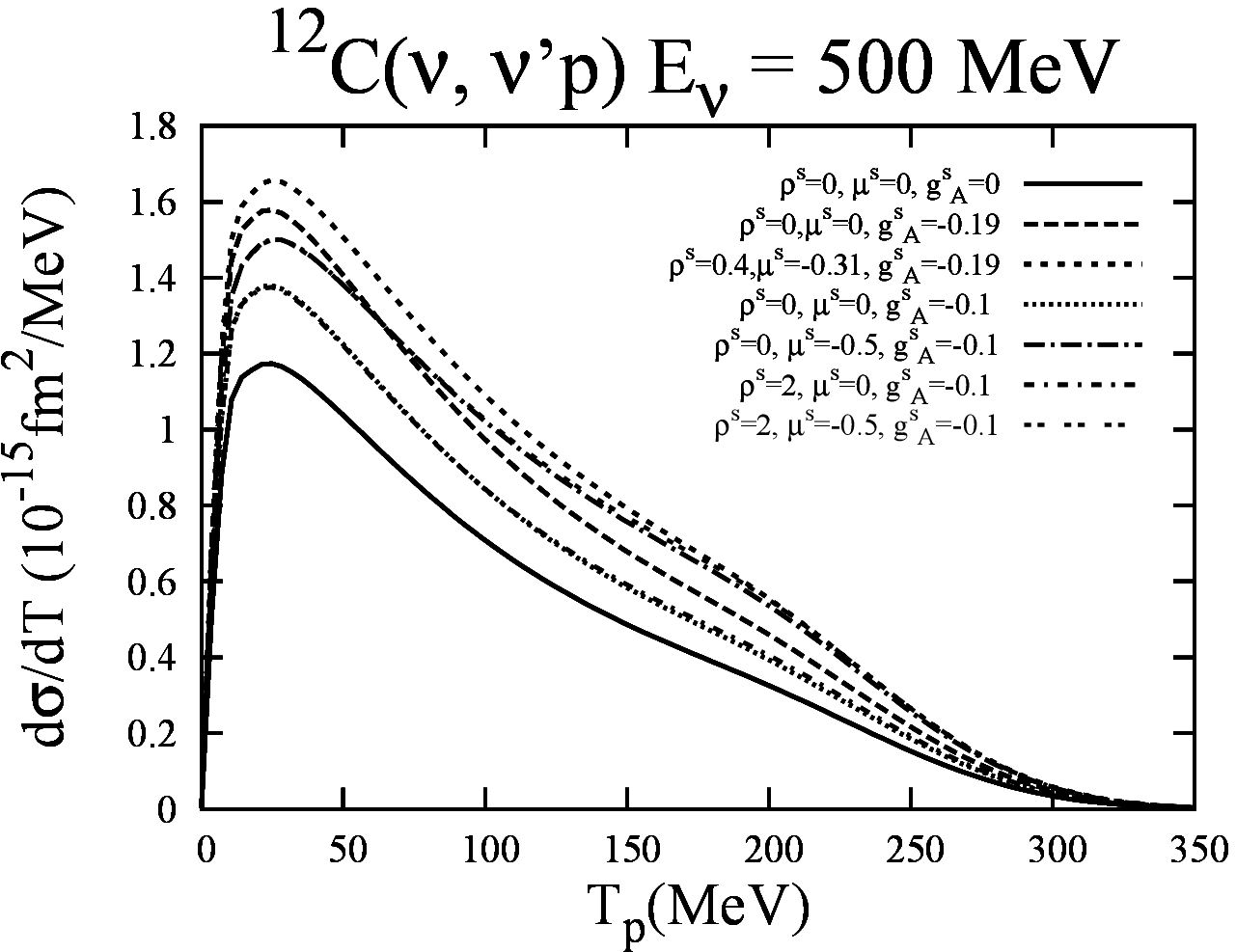}
 \end{center}
 \caption{Same as Figure  \ref{p150} for an incoming neutrino energy equal to 500 MeV.\label{p500}}
\end{figure}

 \begin{figure}
  \begin{center}
  \includegraphics[scale=1.2]{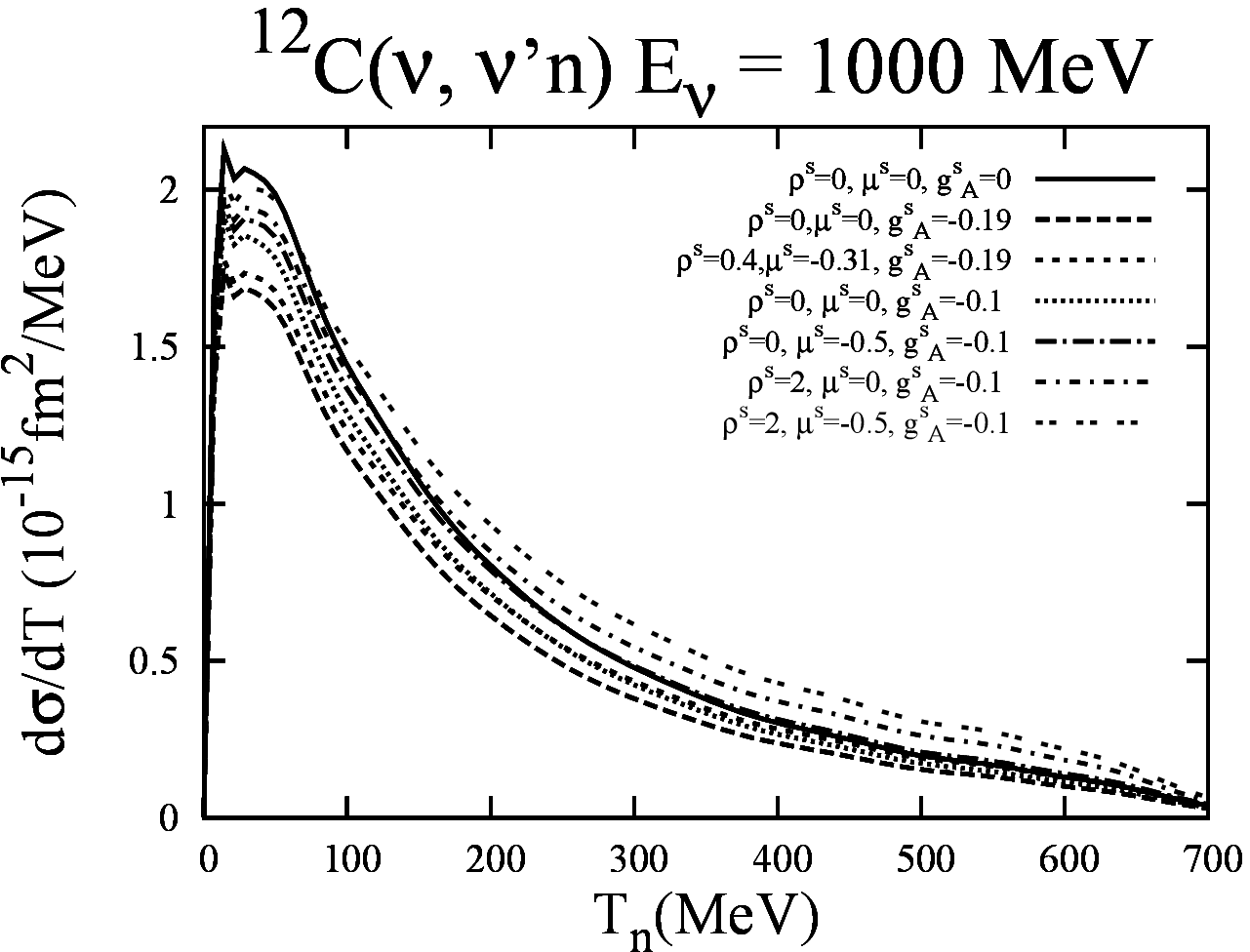}
 \end{center}
 \caption{Same as Figure \ref{n150} for an incoming neutrino energy equal to 1000 MeV.\label{n1000}}
\end{figure}

 \begin{figure}
  \begin{center}
  \includegraphics[scale=1.2]{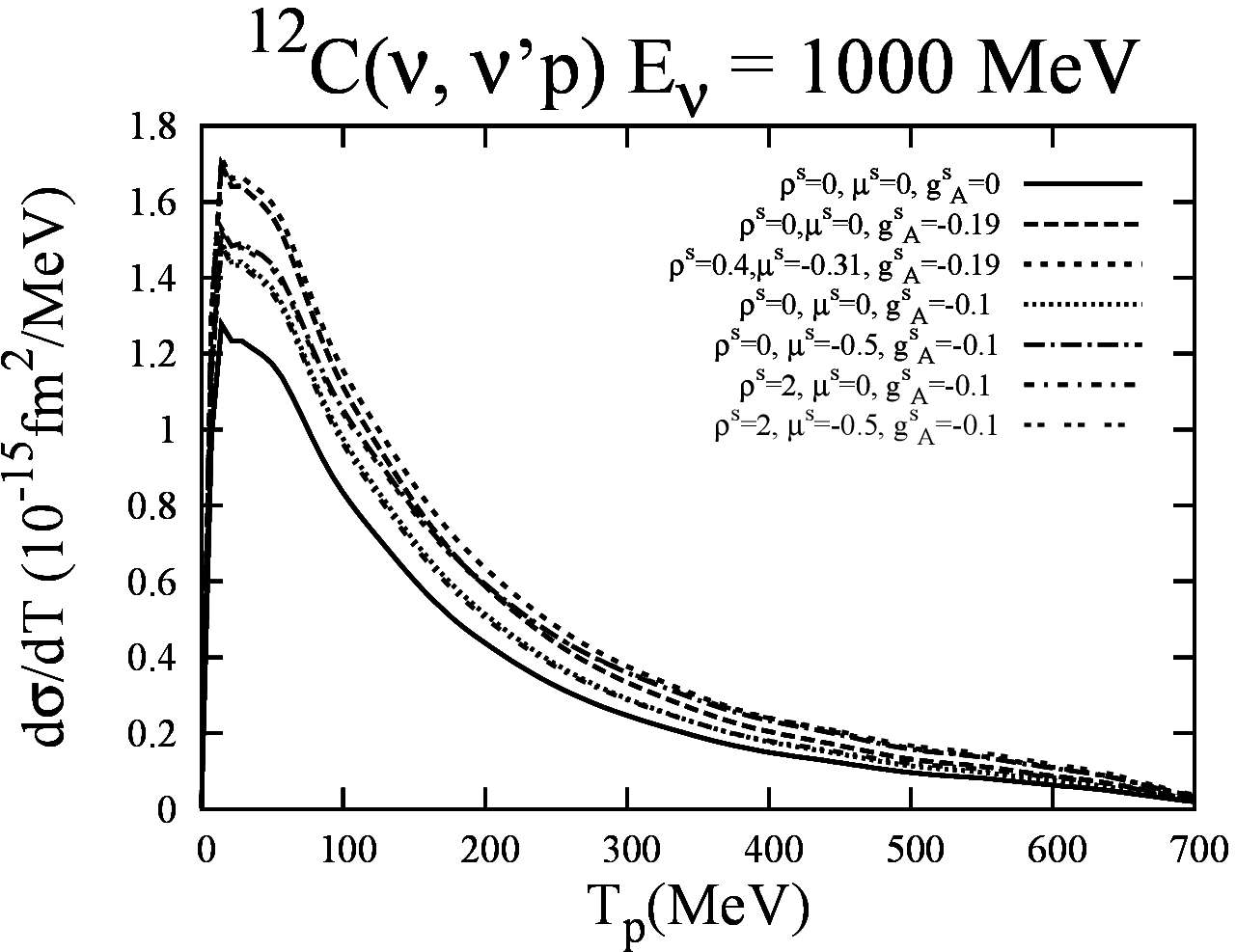}
 \end{center}
 \caption{Same as Figure  \ref{p150} for an incoming neutrino energy equal to 1000 MeV.\label{p1000}}
\end{figure}
 
 \begin{figure}
 \begin{center}
\includegraphics[scale=1.2]{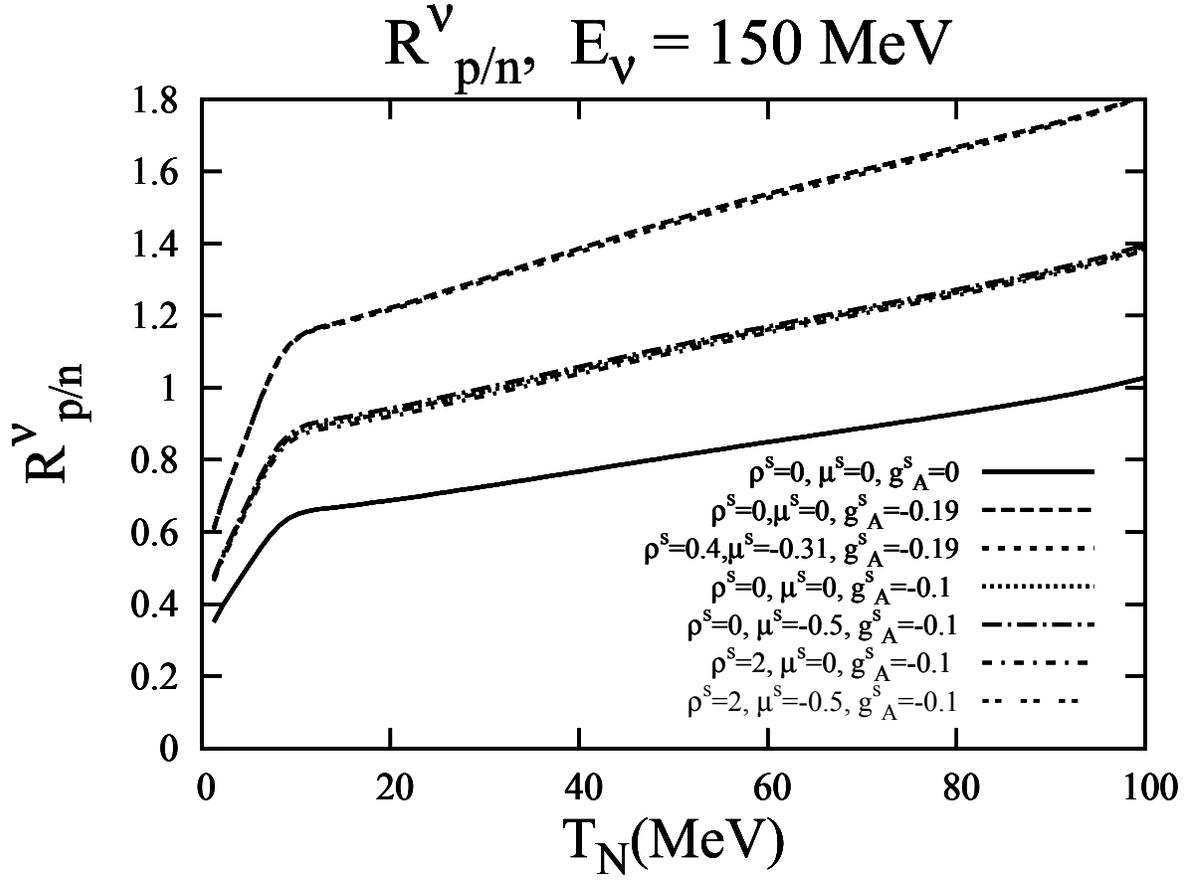}
\end{center}
\caption{Ratio of the proton-to-neutron neutral-current cross sections of the neutrino quasi-elastic scattering on $^{12}$C in terms of the knocked-out nucleon kinetic energy.\label{rpn150}}
\end{figure}

\begin{figure}
  \begin{center}
  \includegraphics[scale=1.2]{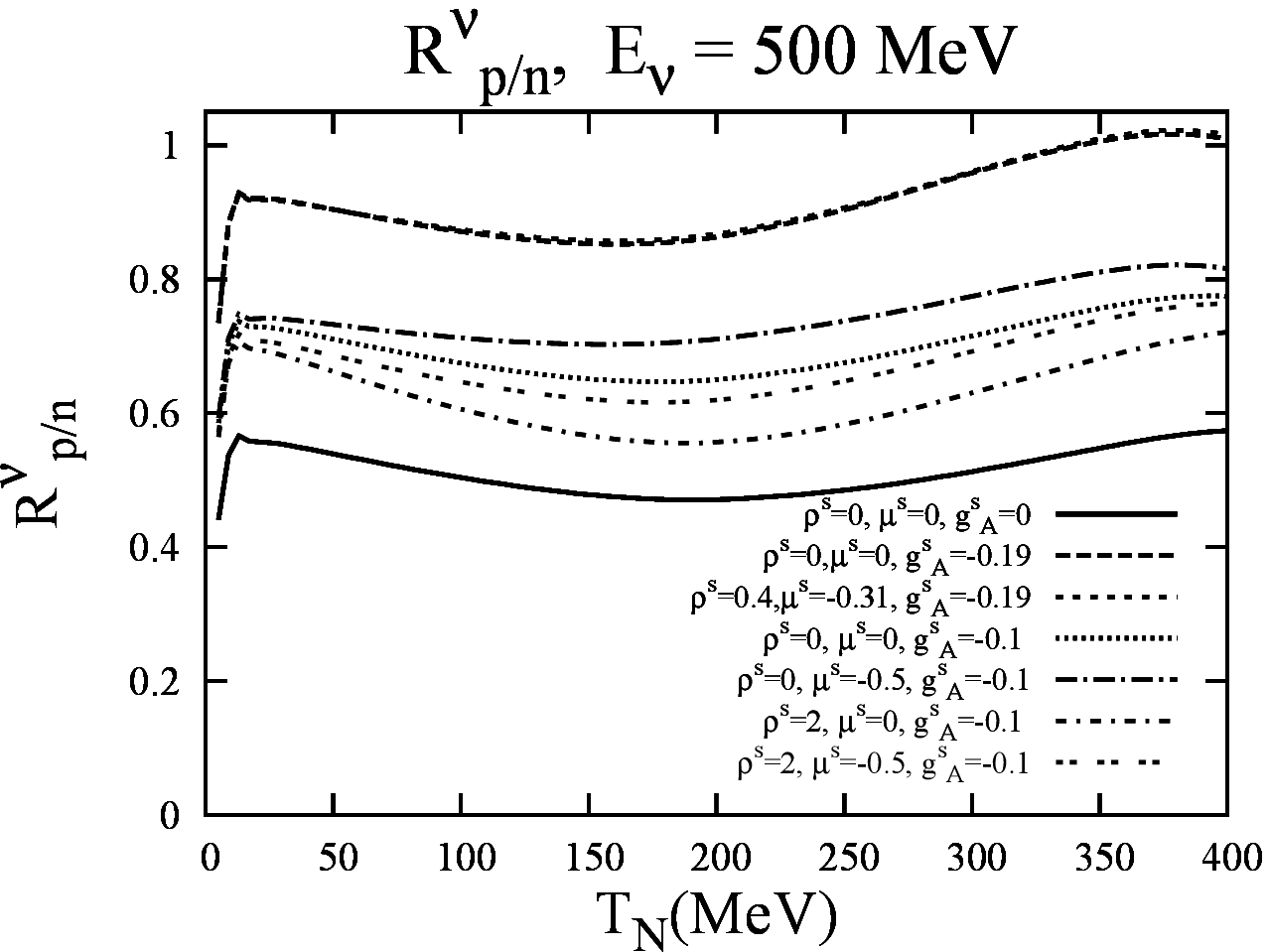}
 \end{center}
 \caption{Same as Figure  \ref{rpn150} for an incoming neutrino energy equal to 500 MeV.\label{rpn500}}
\end{figure}

\begin{figure} 
  \begin{center}
  \includegraphics[scale=1.2]{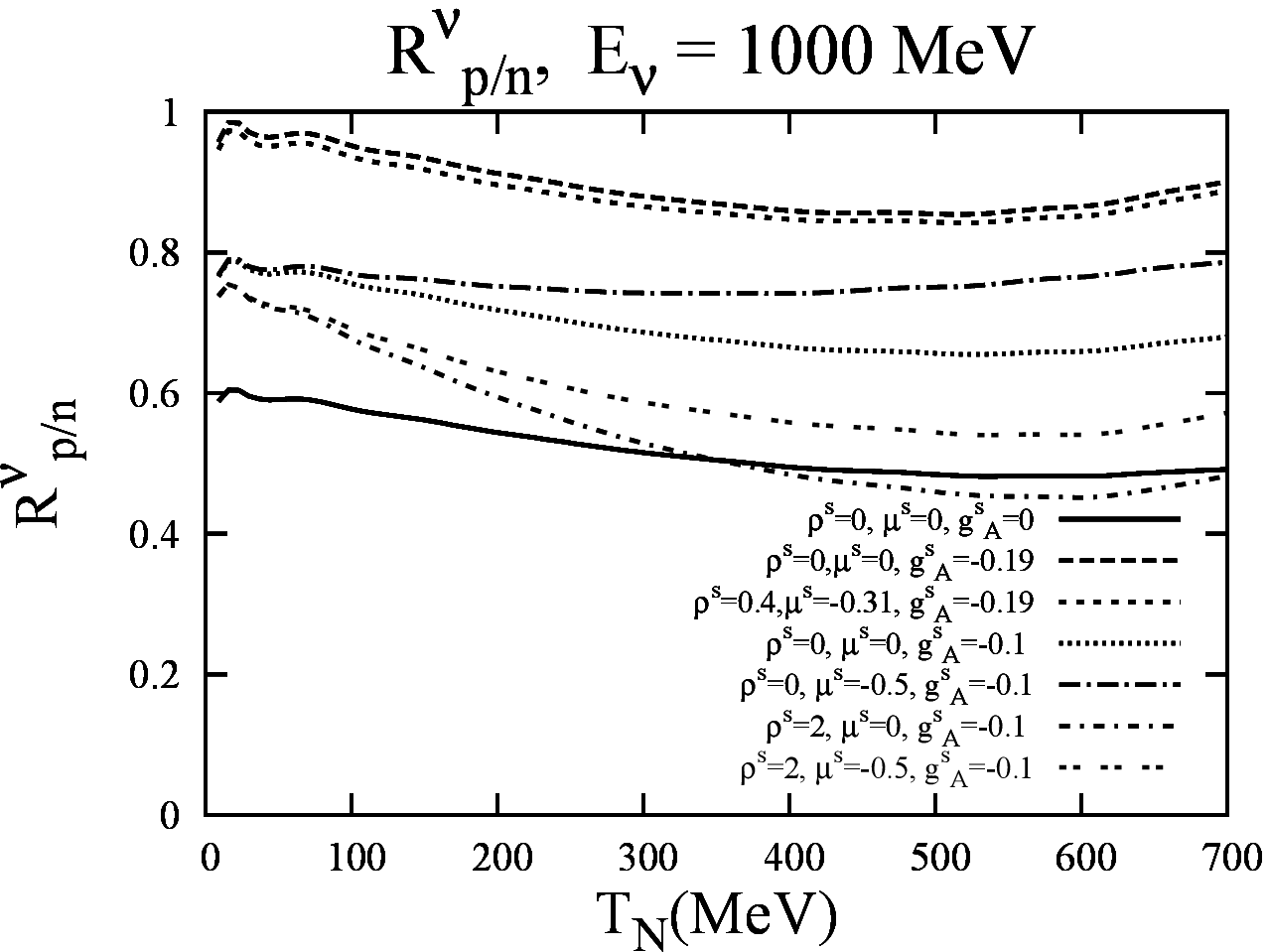}
 \end{center}
 \caption{Same as Figure  \ref{rpn150} for an incoming neutrino energy equal to 1000 MeV.\label{rpn1000}}
\end{figure}

\begin{figure} 
  \begin{center}
  \includegraphics[scale=1.2]{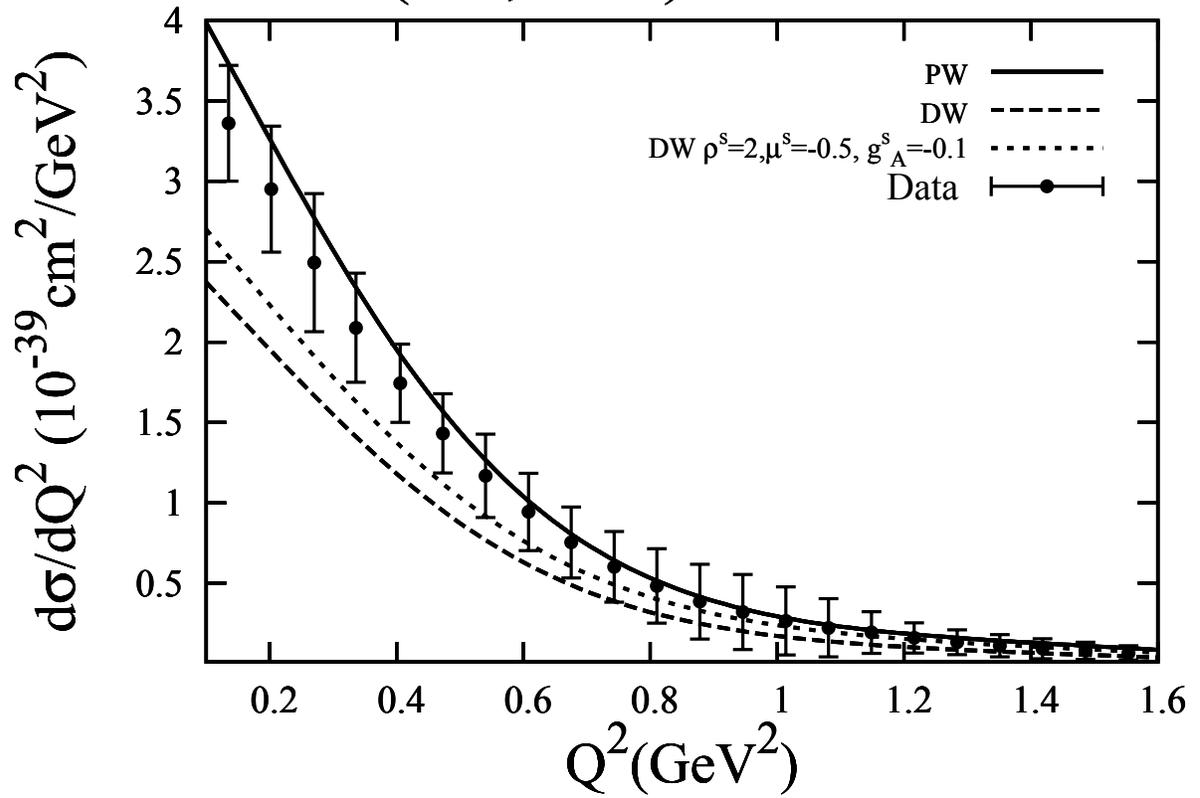}
 \end{center}
 \caption{Neutral-current quasi-elastic neutrino flux-averaged differential cross section scattering on mineral oil (CH$_2$) compared with the MiniBooNE data \cite{Aguilar-Arevalo}.\label{square}}
\end{figure}

\begin{figure} 
  \begin{center}
  \includegraphics[scale=1.2]{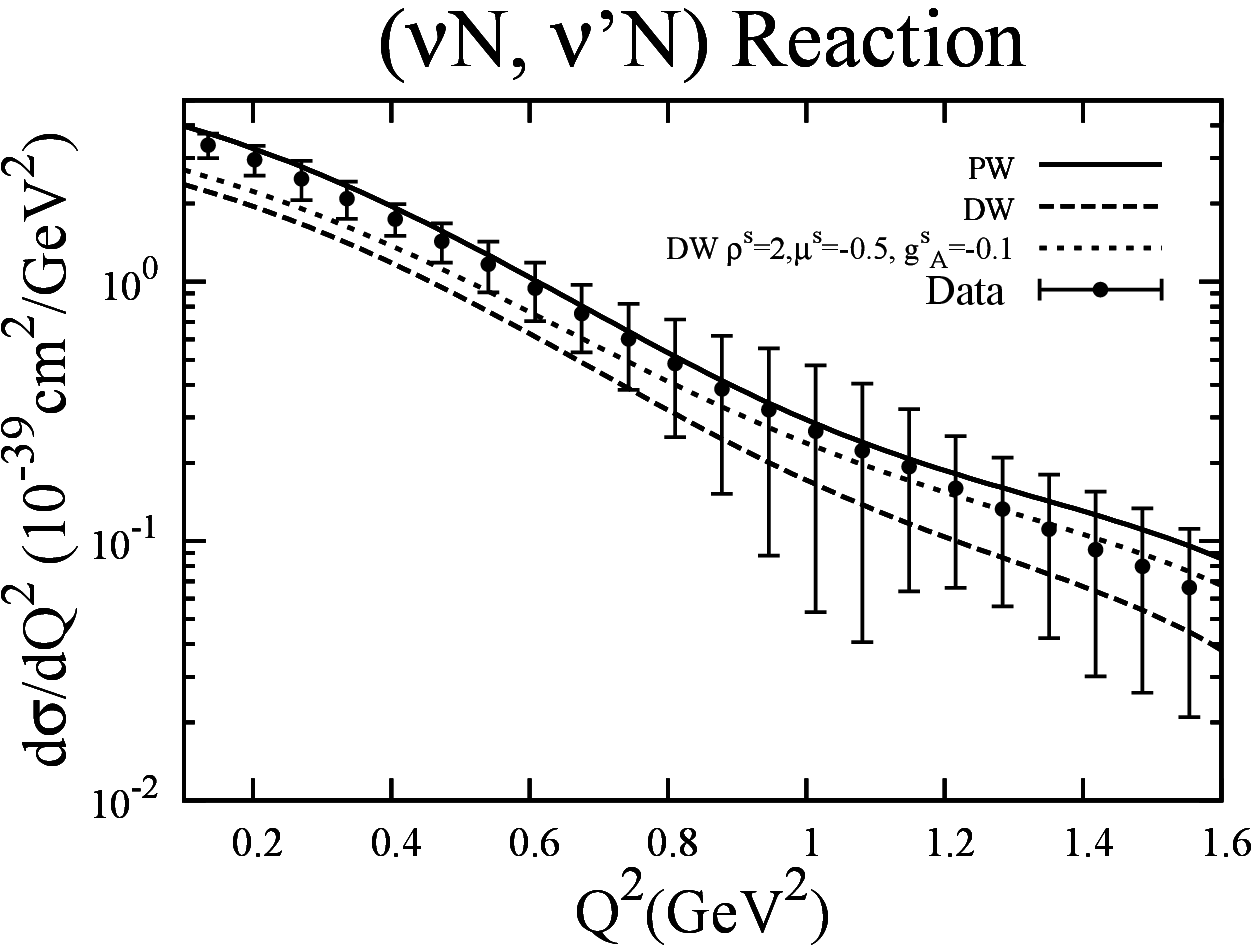}
 \end{center}
 \caption{Neutral-current quasi-elastic neutrino flux-averaged differential cross section represented on a logarithmic scale for the scattering on mineral oil (CH$_2$) compared with the MiniBooNE data \cite{Aguilar-Arevalo}. The log scale displays more clearly the behaviour of the cross sections for large $Q^2$.\label{squarelog}}
\end{figure}

\begin{figure} 
  \begin{center}
  \includegraphics[scale=1.2]{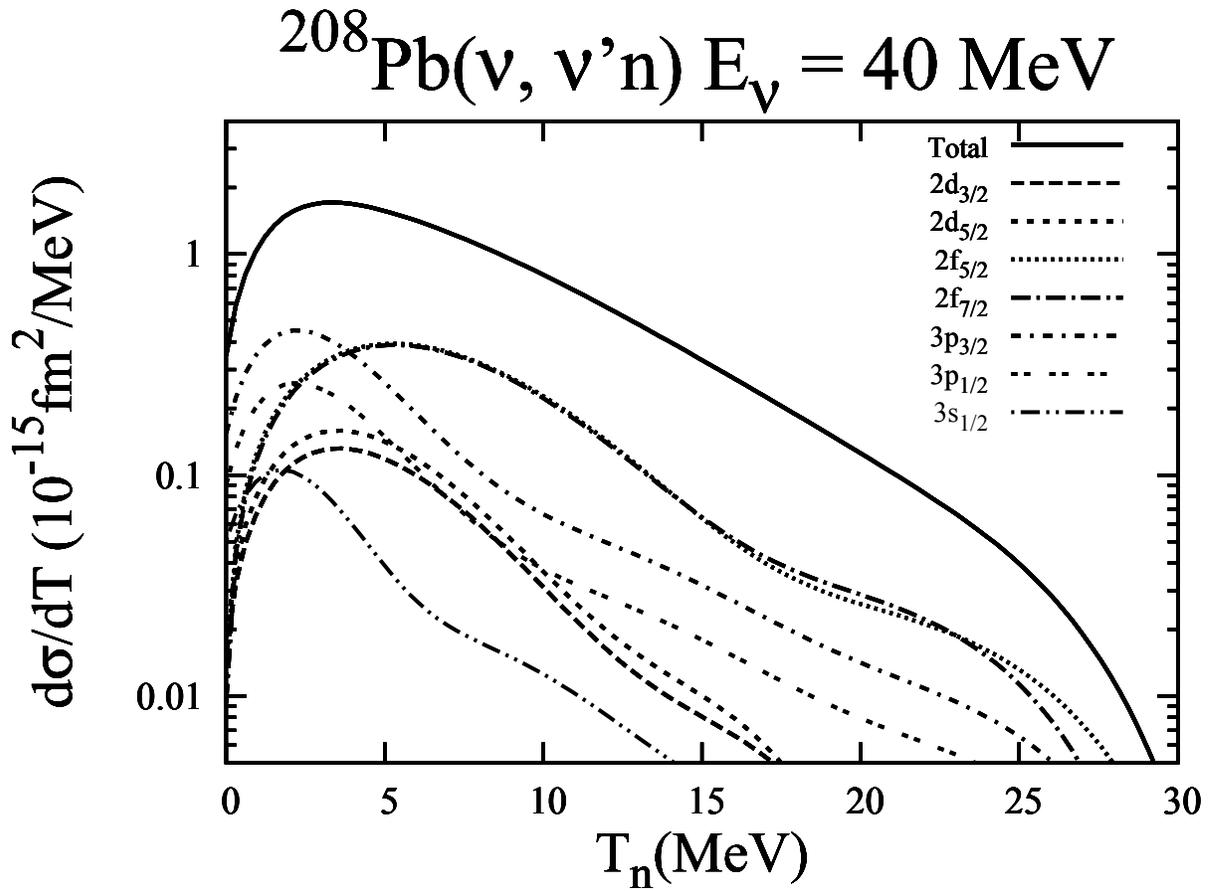}
 \end{center}
 \caption{Differential cross section of the neutral-current neutrino quasi-elastic scattering on $^{208}$Pb in terms of the knocked-out neutron kinetic energy for an incoming neutrino energy equal to 40 MeV. The various lines correspond to the nuclear levels.\label{40MeVAll}}
\end{figure}

\begin{figure} 
  \begin{center}
  \includegraphics[scale=1.2]{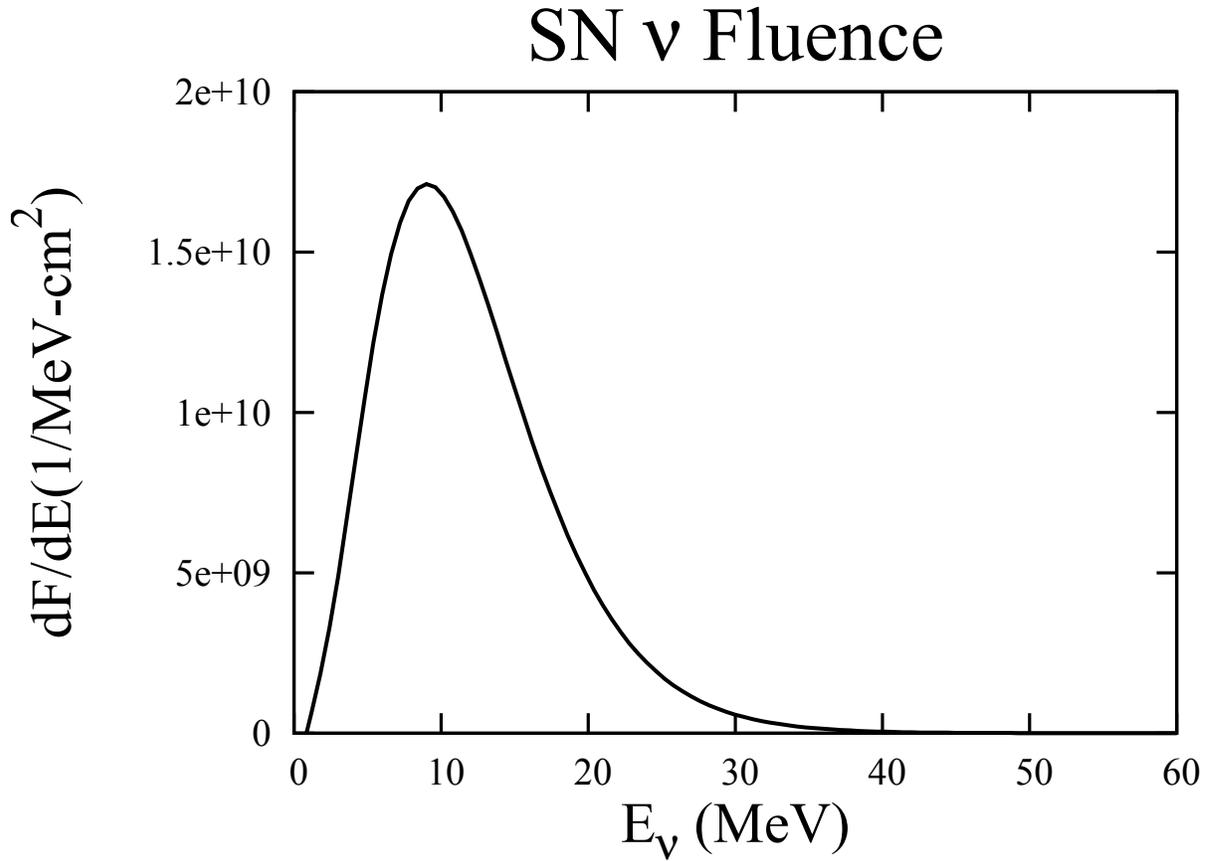}
 \end{center}
 \caption{Fluence distribution for the supernova neutrinos in terms of the emitted neutrino energy. The parameters in Eq.(\ref{fluence}) are taken as $d=1$ (distance of 10 kpc from the emitting supernova to the Earth) and ${\cal E}_\nu=5$ ($5\times10^{52}$ erg).  \label{total1t}}
\end{figure}

\begin{figure} 
  \begin{center}
  \includegraphics[scale=1.2]{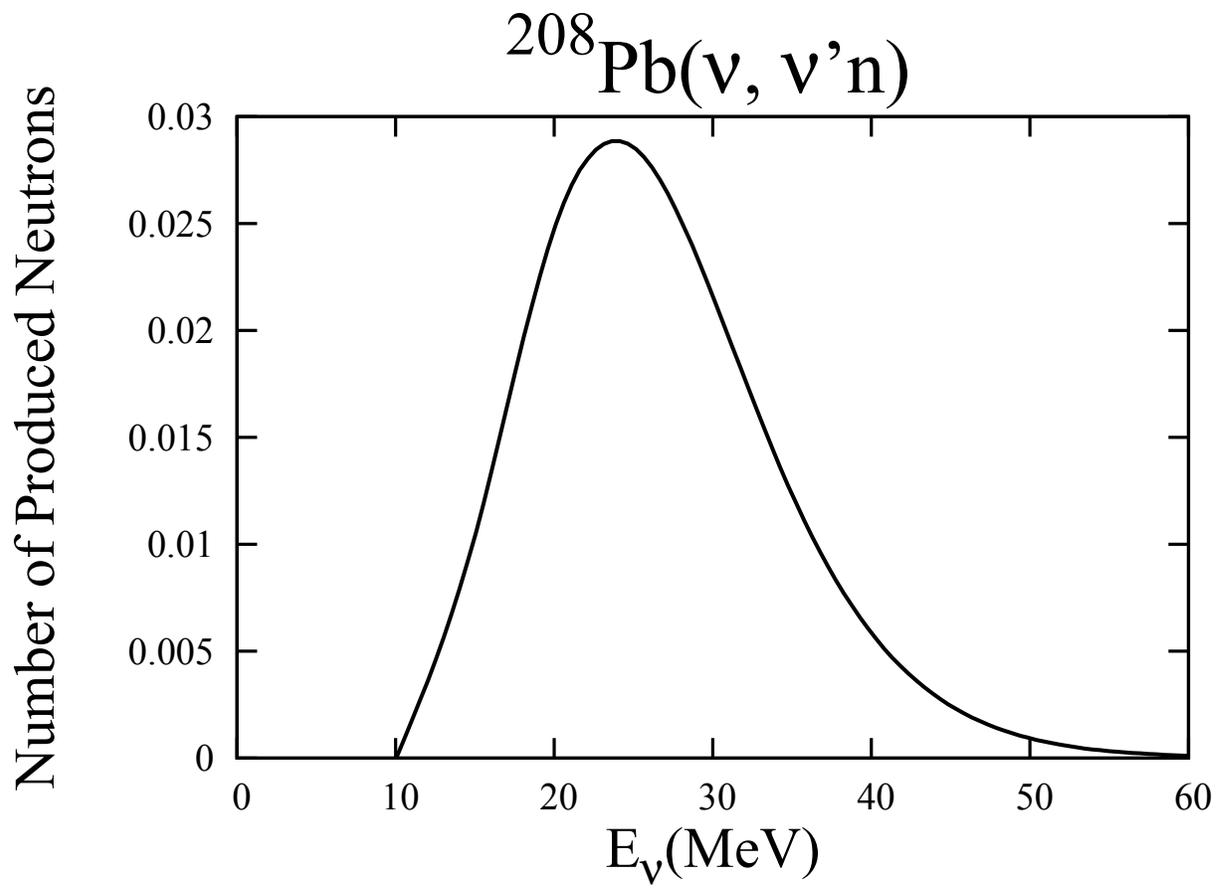}
 \end{center}
 \caption{Number of neutrons produced at HALO-1 in terms of the neutrino energy. \label{event_distribution}}
\end{figure}

\end{document}